\documentclass[
 aip,
 jcp,
 amsmath,amssymb,
 reprint,
 floatfix,
 superscriptaddress
]{revtex4-2}

\usepackage{bm}
\usepackage{booktabs}
\usepackage{graphicx}
\usepackage{tikz}
\usetikzlibrary{arrows.meta,calc,positioning}
\usepackage{hyperref}
\hypersetup{hidelinks}
\usepackage{siunitx}
\usepackage{mathtools}
\usepackage{array}
\usepackage{placeins}

\newcommand{\comm}[2]{\left[#1,#2\right]}

\newcolumntype{L}[1]{>{\raggedright\arraybackslash}p{#1}}

\begin{document}

\title{Compile-once operation graphs for reusable continuous unitary transformations}

\author{Bo Peng}
\email{peng398@pnnl.gov}
\affiliation{Integrated Discovery Sciences Directorate, Pacific Northwest National Laboratory, Richland, Washington 99354, USA}
\author{Niranjan Govind}
\affiliation{Integrated Discovery Sciences Directorate, Pacific Northwest National Laboratory, Richland, Washington 99354, USA}

\date{\today}

\begin{abstract}
Continuous unitary transformations repeatedly evaluate the same operator algebra as the coefficients of a Hamiltonian and its observables evolve. We compile these algebraic relationships once into a numerical operation graph that can be saved and reused for new coefficient sets, additional operators, derivatives, and forward and reverse calculations on CPUs and GPUs. Reuse is demonstrated by applying the same compiled graph to hydrogen and deuterium parameterizations of a reduced molecular model without regenerating the operator equations. The main computational challenge is construction: generated many-body algebra can contain hundreds of millions of contributions before numerical propagation begins. We address this by determining the required storage in advance and writing contributions directly into the final disk-backed representation. In a deliberately large diagnostic-enabled stress test containing more than 350 million generated contributions, peak host memory remains only about 7\% above the final stored size. Cross-checks show that CPU and GPU implementations, forward and reverse operations, and repeated loading of the compiled graph reproduce the tested operations to numerical precision. This compile-once representation separates expensive operator-algebra generation from the numerical transformations that reuse it.
\end{abstract}

\maketitle

\section{Introduction}

\begin{figure*}[t!]
\centering
\includegraphics[width=0.96\textwidth]{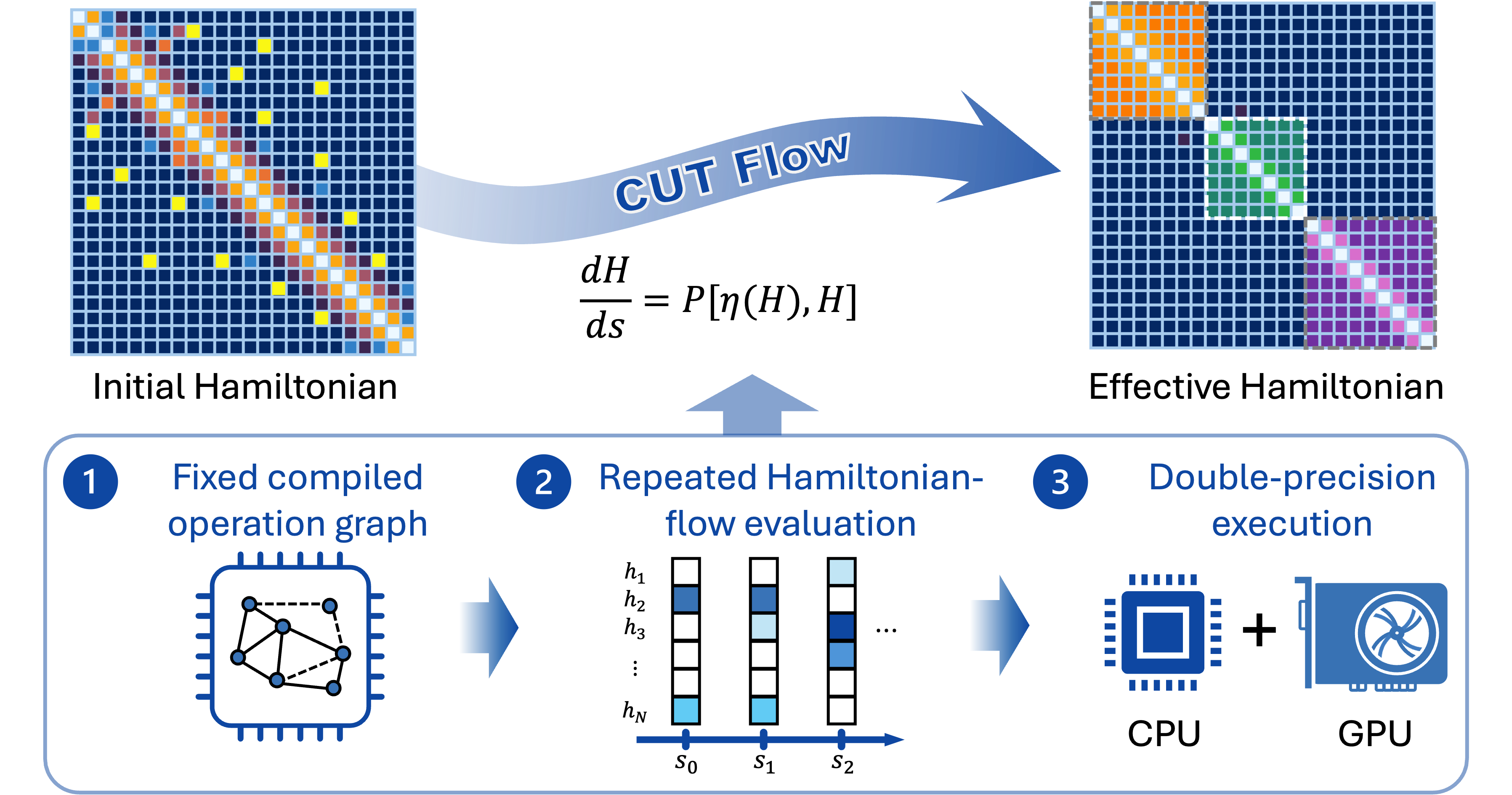}
\caption{\label{fig:workflow}
Compile once, transform many times. A declared operator space and transformation rule are converted once into a numerical operation graph. The stored graph records how operator coefficients contribute to one another and is reused while coefficient vectors, operands, or physical parameters change. The same representation drives Hamiltonian and observable transformations, forward and reverse operations, and CPU or GPU execution. Recompilation is required when the defining algebra changes, not when a supported coefficient set or operand changes.}
\end{figure*}

Transformations of a many-body Hamiltonian are useful only if the resulting representation can be constructed, applied, and reused at a cost that remains manageable as the underlying operator space grows. This issue arises broadly in effective-Hamiltonian theory, where one seeks to weaken couplings between selected and complementary sectors, isolate relevant energy scales, or move Hamiltonians and observables into representations better suited to subsequent calculation. Although the physical objectives differ, these approaches share a useful separation between defining the transformation algebra and applying that transformation to particular Hamiltonians and observables. Once the former has been specified, much of its operator structure can remain fixed while numerical coefficients and transformed quantities change. This fixed-versus-varying structure creates the possibility of algebra reuse.

Historically, the same distinction appears in several forms across the long development of effective-Hamiltonian theory. Projection formalisms separate a target model space from its complement, linked-diagram methods construct effective interactions within that space, and canonical or similarity transformations progressively suppress couplings that are not wanted in the final representation~\cite{BlochHorowitz1958,Feshbach1958,Brandow1967,Kuo1974,EllisOsnes1977,HjorthJensen1995,Suzuki1977,SuzukiOkamoto1984,KuoOsnes1990,Anderson1970,WilsonKogut1974,Wilson1975,GlazekWilson1993,GlazekWilson1994,Perry1994,Cederbaum1989}. Across these formulations, the goal is not simply to simplify a Hamiltonian, but to define a transformation that can also be applied consistently to the observables and auxiliary operators needed in the transformed representation.

Among these approaches, continuous unitary transformations (CUTs) realize the effective-Hamiltonian strategy through a continuous differential flow.~\cite{Wegner1994,Mielke1998,Lenz2001,Kehrein2006,Stauber2003}. For a flow coordinate $s$, the Hamiltonian evolves according to
\begin{equation}
\frac{dH(s)}{ds}=\comm{\eta(s)}{H(s)},\qquad
\eta^\dagger(s)=-\eta(s),
\label{eq:flow-intro}
\end{equation}
where the generator $\eta(s)$ determines which matrix elements, couplings, or blocks are progressively removed. The same accumulated transformation must be applied to every observable or auxiliary operator,
\begin{equation}
\frac{dO(s)}{ds}=\comm{\eta(s)}{O(s)},
\label{eq:observable-flow-intro}
\end{equation}
because transforming the Hamiltonian alone would leave measured quantities in a different representation. Consistent operator evolution is therefore an intrinsic part of the transformation~\cite{Hackl2008,Anderson2010,Schuster2014,Parzuchowski2017,Thomson2024}. CUTs and closely related similarity flows have consequently been developed for lattice models, electron--phonon and spin--boson systems, nuclear Hamiltonians, quantum chemistry, nonequilibrium dynamics, and other many-body settings~\cite{Mielke1997,Kehrein1995,KehreinMielke1997,KehreinMielke1998,Knetter2000,Krull2012,Schamriss2024,BognerKuoSchwenk2003,BognerFurnstahlPerry2007,Jurgenson2009,Tsukiyama2011,Hergert2016,Stroberg2019,Morris2015,Watson2016,ThomsonSchiro2023,Liu2026VCUT}.

Electron–vibrational Hamiltonians provide a particularly transparent setting in which the reusable structure of CUTs becomes visible. Electronic excitation changes the forces experienced by the nuclei, while nuclear motion can in turn redirect electronic populations and coherences. Such feedback underlies proton transfer, charge and energy transfer, photochemical switching, and other nonadiabatic processes~\cite{Demchenko2013,Kumpulainen2017,Sedgwick2018,Zhou2018}. In a second-quantized representation, however, changing vibrational frequencies, couplings, isotope masses, or observable coefficients does not necessarily change the multiplication rules of the underlying electronic and bosonic operators. The excited-state proton-transfer molecule 10-hydroxybenzo[$h$]quinoline (HBQ) offers a concrete example: its electronic dynamics are coupled to proton and skeletal vibrations, while isotope substitution changes physical parameters without changing the basic operator structure~\cite{Takeuchi2005,Higashi2011,Lee2013,Kim2020,Biswas2026,Picconi2021,Zhang2023,Zhang2025,LeDe2025,Loe2021,Nimmrich2024}. We use a reduced HBQ model below to demonstrate this coefficient-reuse property: the same compiled algebraic representation can accept distinct H and D parameter sets without regenerating the operator equations. This example also highlights an important distinction that applies more generally to finite CUT calculations. Reusing a fixed algebraic representation is a computational property; whether the retained operator space provides an adequate physical approximation is a separate question.

A finite implementation of a CUT therefore presents two conceptually distinct challenges. The first is physical. Exact commutators generally generate operators outside any finite retained space, often with increasing particle rank or bosonic degree, so closing the flow requires projection or truncation. Convergence of the numerical integrator therefore does not by itself establish convergence of populations, coordinates, coherences, or other observables; the adequacy of the retained operator space must ultimately be tested against the quantity of interest~\cite{Drescher2010,Savitz2017,Gong2020,Etienney2026}. The second challenge is computational and remains even after that finite algebra has been chosen. Expanding its commutators into coefficient equations can generate a very large number of contributions with strongly nonuniform structure: one operator pair may commute, another may generate a single destination, and another may produce many terms after exact bosonic normal ordering. Materializing every contribution first as a separate symbolic or programming-language object can therefore require far more transient memory than the compact numerical representation ultimately needed for execution. It is this second problem, together with the opportunity to reuse the resulting algebra across many numerical calculations, that motivates the present work.

Substantial prior work addresses neighboring parts of this computational pipeline. Tensor Contraction Engine and related many-body code-generation systems move symbolic derivation outside the production calculation and generate executable tensor expressions or contraction programs~\cite{Hirata2003,Mutlu2023,Brandejs2025,DePrince2025Pdaggerq}. More recent algebraic tools automate second-quantized manipulations and block-diagonalization calculations~\cite{Pymablock2025}. Perturbative and finite-basis CUT approaches precompute the commutator coefficients that define flow equations within selected operator spaces~\cite{Knetter2000,Krull2012,Schamriss2024}, while the Magnus formulation of the in-medium similarity renormalization group retains an accumulated transformation that can subsequently be applied to additional operators~\cite{Morris2015}. Sparse-matrix and accelerator libraries, in turn, provide mature machinery for executing large numerical kernels~\cite{Gustavson1978,Fales2015,Bell2009,Deveci2018}. These developments establish symbolic equation generation, finite-basis flow construction, transformation reuse, and high-performance execution as powerful ideas individually. The opportunity pursued here is to make the generated CUT algebra itself a durable numerical object that can be reused across numerical operations, physical coefficient sets, and processor backends.

We therefore treat the coefficient equations produced by a finite CUT as a computational object that can be compiled independently of the numerical trajectory that will later consume them. We call this representation an \emph{operation graph}. Each stored contribution identifies the input coefficients involved in one algebraic product, its numerical factor, and the output coefficient to which it contributes. Once these relationships have been generated, subsequent evaluations require only changing coefficient vectors or operands; the underlying operator multiplication rules need not be rediscovered. Finite-basis CUT formulations already exploit the fact that these commutator coefficients are fixed. The distinction here is to promote that fixed algebra to a persistent runtime object whose stored records are shared by forward, adjoint, derivative, and heterogeneous-hardware execution. The operation graph stores these relationships as persistent numerical records consumed by shared execution routines, rather than only as generated source code. It represents the CUT vector field, not the accumulated transformation of a completed flow or one fixed numerical linear map. Numerical coefficients enter only when the graph is executed, allowing the same stored structure to provide Hamiltonian and observable propagation, forward and adjoint operations, derivatives, and CPU or GPU execution. In this sense, the graph is not simply a compressed record of a symbolic calculation, but a persistent numerical encoding of the transformation algebra.

The size of this compiled representation creates its own systems problem. A graph containing hundreds of millions of generated contributions should not require another graph-sized collection of temporary symbolic objects merely to construct it. We therefore determine the required storage before writing contributions directly into the final disk-backed arrays. Once constructed, the graph can be checked, saved, and reloaded in another process or compute allocation without repeating the operator-algebra generation. This separates a potentially expensive one-time compilation stage from the many numerical calculations that reuse its result.

Figure~\ref{fig:workflow} summarizes this separation between algebra construction and repeated execution. We demonstrate it at several levels. First, a reduced HBQ example shows that one compiled graph can be reused for distinct H and D coefficient sets, providing a physical instance of parameter reuse. Second, an intentionally large Pauli--boson construction test asks whether hundreds of millions of generated contributions can be compiled without a comparable temporary-memory expansion. Third, independent CPU and GPU implementations exercise forward, reverse, and derivative traversals of the graph. Finally, we resolve compilation, reload, setup, and execution costs separately to identify when persistence is advantageous and when a compact matrix or same-process representation remains preferable. Together, these tests establish a compile-once execution layer for generated CUT algebra while keeping the accuracy of the chosen finite operator space as a separate physical question.


\section{Method: compile once, transform many times}
\label{sec:method}

The implementation follows the separation introduced above between defining an operator transformation and repeatedly applying it. We organize the method accordingly. First, the physical model, generator, and finite operator registry specify the projected CUT equations that are to be solved. Second, the resulting operator products are compiled into a coefficient-level numerical map whose structure no longer depends on a particular coefficient vector. Third, this map is constructed and persisted without first retaining a second graph-sized symbolic representation. Finally, the graph supports forward and adjoint propagation, derivatives, and multiple transformed operators. These four steps define, respectively, what is compiled, how it is encoded, how it is built, and how it is reused.

\subsection{Projected CUT and finite Pauli--boson registry}

The first step is physical rather than computational: we specify the finite operator algebra whose equations the compiler will reproduce. For a flow coordinate $s$, the Hamiltonian and any coflowed operator $X$ obey
\begin{equation}
\dot H=\comm{\eta(H)}{H},\qquad
\dot X=\comm{\eta(H)}{X},
\label{eq:coflow}
\end{equation}
where the block generator
\begin{equation}
\eta=\comm{H_{\mathrm d}}{H_{\mathrm{od}}}
\label{eq:block-generator}
\end{equation}
separates block-diagonal and off-diagonal coordinates
\begin{equation}
H_{\mathrm d}=\sum_B P_BHP_B,\quad
H_{\mathrm{od}}=H-H_{\mathrm d}.
\end{equation}
The projectors $P_B$ define the desired block structure and are physical inputs rather than compiler choices. The reported workloads use the joint electronic-$\sigma_z$/boson-number basis to define the diagonal sector, so the implemented coefficient mask retains the $I$ and $\sigma_z$ monomials with $\bm c=\bm a$ within dagger-closed Hermitian registries with nonempty diagonal and off-diagonal sectors.

For $M$ boson modes, creation and annihilation multi-indices $\bm c,\bm a\in\mathbb N_0^M$, and a Pauli label $\tau\in\{0,x,y,z\}$, we write the canonical operator labels as
\begin{equation}
B_{\tau,\bm c,\bm a}=\sigma_\tau
\prod_{m=1}^{M}(b_m^\dagger)^{c_m}b_m^{a_m}.
\label{eq:monomial}
\end{equation}
The complete degree-$p$ registry contains every label satisfying $|\bm c|+|\bm a|\le p$ and has dimension
\begin{equation}
N_{\mathrm{op}}(M,p)=4\binom{2M+p}{p}.
\label{eq:basis-size}
\end{equation}
Products are evaluated in the full bosonic algebra and normal ordered before membership in the finite registry is tested. This ordering is essential: contractions of products whose intermediate degree exceeds $p$ can still generate lower-degree terms that belong to the retained space.

Let $P$ denote projection into the chosen registry. For any coflowed operator,
\begin{equation}
F_X=P\comm{\eta(H)}{X},\qquad
C_X=(\mathcal I-P)\comm{\eta(H)}{X}.
\label{eq:projected-flow}
\end{equation}
$F_X$ defines the equations actually propagated within the selected operator space. $C_X$ contains terms generated immediately outside that space. These outside-space terms do not feed back into the selected flow and need not be stored for production calculations; when retained, they provide an algebraic diagnostic rather than an observable-error certificate.

Once the registry, generator, normal-ordering convention, and projection have been fixed, the topology of the coefficient equations is fixed as well. Numerical values of $H$ or $X$ change during the flow, but the operator multiplication rules connecting their coefficients do not. This provides the compilation boundary used in the next step.

\subsection{From operator products to a stored numerical map}

\begin{figure*}[t!]
\centering
\hspace*{-10mm}
\resizebox{1.05\textwidth}{!}{\input{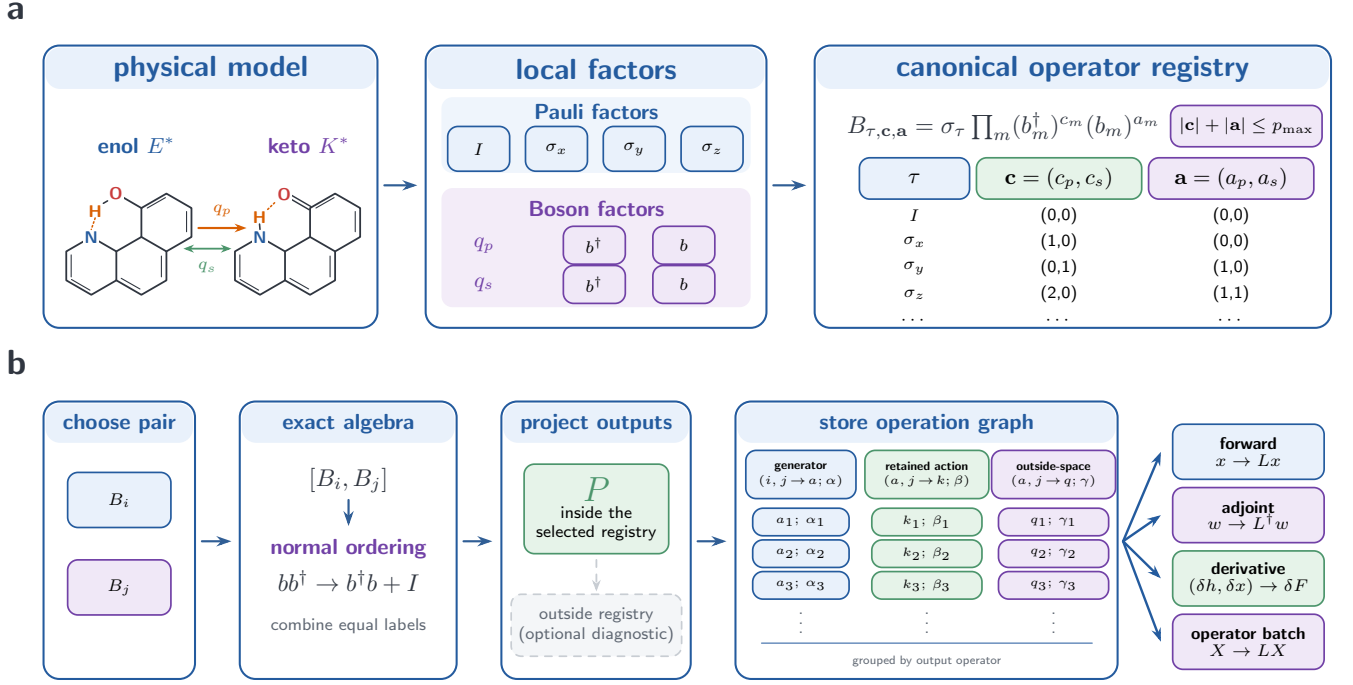}}
\caption{\label{fig:method-map}
Compilation of Pauli--boson algebra into a reusable numerical operation graph. A physical electronic--vibrational model is first expressed through local Pauli and boson factors and mapped to a canonical finite operator registry. Exact operator products are evaluated and normal ordered before projection. Contributions inside the registry define the production equations, whereas generated terms outside the registry may be retained separately for diagnostics. Each surviving algebraic term is then stored as indexed input and output coordinates together with a numerical factor. Grouping records by output yields the numerical map used by the execution routines.}
\end{figure*}

We therefore compile the projected algebra independently of any particular trajectory. Figure~\ref{fig:method-map} illustrates this mapping from local electronic and bosonic factors to canonical operator labels, exact products, and finally numerical contribution records.

Let $h$ and $x$ denote the coefficient vectors of $H$ and $X$, respectively, and let $D$ select the block-diagonal Hamiltonian coordinates so that $Dh$ and $(I-D)h$ represent the two factors entering the generator. For all reported workloads, the generator registry enumerates every nonzero term of the full commutator in Eq.~\eqref{eq:block-generator} and therefore represents $\eta$ exactly without an additional generator truncation, with $P$ applied only after the subsequent commutator has been normal ordered. Each generated term can then be reduced to a small set of integer indices and one algebraic factor. The stored generator and its retained and outside-space actions take the form
\begin{equation}
\left\{
\begin{aligned}
\xi_a(h) &= \sum_{e\in\mathcal E_\eta(a)}
\alpha_e(Dh)_{i_e}[(I-D)h]_{j_e},\\
F_k(h,x) &= \sum_{e\in\mathcal E_L(k)}
\beta_e\xi_{a_e}(h)x_{j_e},\\
C_q(h,x) &= \sum_{e\in\mathcal E_C(q)}
\gamma_e\xi_{a_e}(h)x_{j_e}.
\end{aligned}
\right. \label{eq:graph-actions}
\end{equation}
Here $\mathcal E_\eta$, $\mathcal E_L$, and $\mathcal E_C$ denote the stored contributions to the generator, retained action, and optional outside-space action. An individual record specifies the relevant input coordinates, its output coordinate, and the precomputed algebraic factor $\alpha_e$, $\beta_e$, or $\gamma_e$.

Records that contribute to the same output are stored together. Evaluation then consists of multiplying the current coefficient values by the stored factors and reducing those contributions directly into the appropriate output coordinate. No dense matrix representing the complete coefficient-space action needs to be formed. Unlike a sparse matrix representing one fixed linear operator, the graph retains the factorized dependence of the CUT action on the evolving Hamiltonian coefficients $h$. The numerical map $L(h)$, its adjoint, and its directional derivatives are therefore instantiated from the same stored algebra as $h$ changes during the flow. This indexed form is the coefficient-level realization of the operation graph introduced above.

The operation graph specifies \emph{what} numerical information must be stored, but this alone does not determine whether constructing that representation is practical at large scale. A graph with hundreds of millions of contributions can lose its memory advantage if an equally large collection of temporary symbolic records must first be materialized. This motivates a construction procedure designed around the final storage itself.

\subsection{Bounded-memory construction and persistence}

A straightforward compiler can generate every symbolic contribution in memory and only afterward pack those records into numerical arrays. We avoid this graph-sized intermediate through two deterministic traversals of the same algebra.

The first traversal determines how many contributions will be written to each output and therefore fixes the lengths of the final arrays. The compiler can then allocate disk-backed index and factor arrays at exactly those sizes. A second traversal repeats the same algebraic generation and writes each contribution directly into its final location. At any stage, the compiler therefore retains the finite operator bases, counting information, and the current local commutator expansion rather than another complete in-memory copy of the graph.

Persistence is built into this design rather than added after compilation. The saved bundle records the operator registry, block mask, ordering and projection rules, factor precision, and identifiers needed to establish which algebra the numerical arrays represent. A completed graph can consequently be checked and reloaded in a different process or compute allocation without repeating normal ordering or symbolic equation generation.

Constructing and saving the graph pays the algebraic cost once. Its utility, however, depends on whether one compiled graph can serve the different numerical operations required after compilation. We therefore design the execution layer so that forward propagation, reverse accumulation, derivatives, and batches of transformed operators are different traversals of the same underlying records rather than separately generated equations.

\subsection{Forward, adjoint, derivative, and multi-operator reuse}

At a given flow coordinate, let the current Hamiltonian coefficients $h(s)$ define the linear action of the projected CUT on an operand,
\begin{equation}
L(h)x \equiv F(h,x).
\end{equation}
The forward coflow is then
\begin{equation}
\dot x=L(h)x.
\label{eq:forward-coeff}
\end{equation}

To define the corresponding reverse action, we introduce a dual coefficient vector $w$ paired with the forward vector $x$. The vector $w$ is an adjoint computational variable rather than, in general, a new physical observable. It obeys
\begin{equation}
\dot w=-L(h)^\dagger w.
\label{eq:dual}
\end{equation}
The minus sign and adjoint ensure that, when the forward and dual coefficient spaces coincide,
\begin{equation}
\frac{d}{ds}\bigl(w^\dagger x\bigr)=0.
\label{eq:duality-invariant}
\end{equation}
For a given Hamiltonian trajectory $h(s)$, this dual is paired with the linear operand equation. Because perturbing the Hamiltonian also changes its generator, Hamiltonian-direction derivatives additionally include the induced generator variation in Eq.~\eqref{eq:directional-derivative}. The invariant provides a trajectory-level consistency check of the forward and reverse implementations.

At the graph level, the distinction between the two propagation directions is simple. Forward execution accumulates each stored contribution into its output coordinate. The adjoint traversal uses the same algebraic factor but accumulates back into the associated input coordinate with the required conjugation. Thus reverse propagation does not require a separately generated set of operator equations.

This bidirectional use of the graph extends naturally to derivatives. Because each graph record retains the elementary coefficient products from which $F(h,x)$ is assembled, perturbations of either the Hamiltonian coefficients or the operand coefficients can be differentiated directly through the stored representation. For infinitesimal changes $\delta h$ and $\delta x$, the corresponding variation of the projected action is
\begin{equation}
 \delta F
 =
 \frac{\partial F}{\partial h}\,\delta h
 + L(h)\,\delta x .
 \label{eq:directional-derivative}
\end{equation}
The first term describes how changing the Hamiltonian coefficients modifies the generator and hence the projected action, whereas the second is the linear response to changing the operand itself. Both terms are evaluated by applying product rules to those stored contributions, without forming a dense Jacobian. The corresponding adjoint derivative operations similarly reuse these factors to propagate sensitivities backward.

Finally, the projected-action graph can accept many operand vectors at once. Once a transformation has been completed for a library of source operators, their transformed coefficient vectors can be reused to form new operator combinations without another symbolic expansion or another compilation of the underlying algebra. Forward propagation, adjoint propagation, derivatives, and multi-operator evaluation are therefore different numerical consumers of one authoritative compiled representation. This distinction between one-time algebra generation and repeated numerical use is the basis of the benchmarks that follow.


\section{Results}\label{sec:results}

The compile-once construction leads to four practical questions. Can the same compiled algebra be reused when physical parameters change? Can a very large generated graph be constructed without a graph-sized temporary representation? Can forward, adjoint, derivative, CPU, and GPU calculations all consume the same stored algebra consistently? And, once the graph exists, when does persistence and repeated reuse become worthwhile? We address these questions in turn.

\subsection{One compiled graph can be reused across physical parameter sets}

Isotope substitution provides a concrete physical test. In the published one-proton-mode HBQ model~\cite{Zhang2023}, replacing hydrogen by deuterium changes the numerical Hamiltonian and preparation parameters but leaves the one-mode Pauli--boson labels and their multiplication rules unchanged. The associated operation graph can therefore be compiled once and evaluated with either physical parameter set.

\begin{figure*}[t!]
 \centering
 \includegraphics[width=0.96\textwidth]{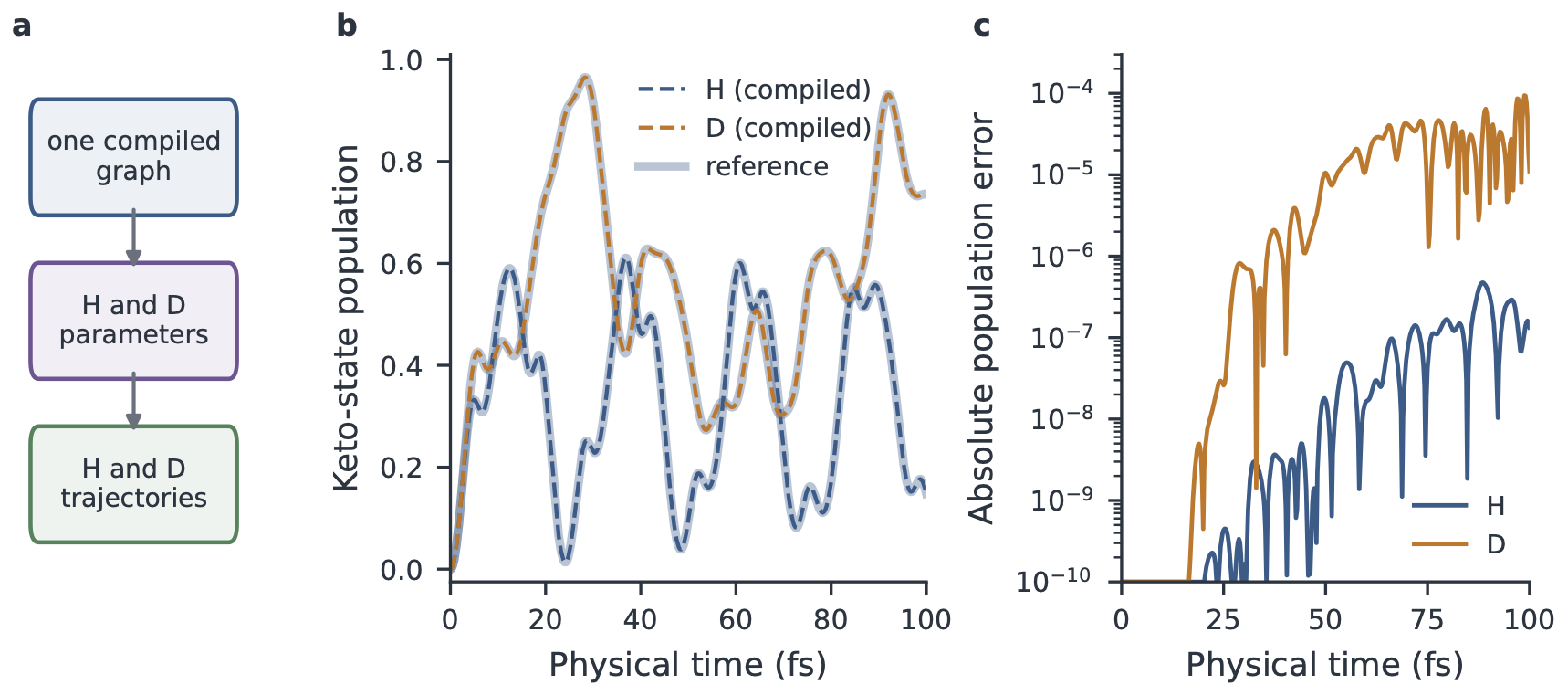}
 \caption{\label{fig:isotope-reuse}
 Reuse of one compiled operation graph for H and D parameterizations of the one-mode HBQ model. (a) The operator graph is unchanged while the physical parameters are replaced. (b) Keto-state populations obtained from the compiled graph are compared with independently converged direct finite-Fock references over 0--100~fs. (c) Absolute population errors, $\left|P_K^{\mathrm{graph}}(t)-P_K^{\mathrm{ref}}(t)\right|$, remain below $4.736\times10^{-7}$ for H and $9.405\times10^{-5}$ for D in this fixed control. The compiled graph uses the complete one-mode registry through total bosonic degree 24 (1,300 coordinates).}
\end{figure*}

Figure~\ref{fig:isotope-reuse} makes the distinction between fixed algebra and changing parameters explicit. The same compiled graph is used for both isotopologues; only the numerical coefficients change. Against independently converged finite-Fock references over 0--100~fs, the maximum absolute errors in the keto-state population are $4.736\times10^{-7}$ for H and $9.405\times10^{-5}$ for D, with RMS errors of $1.107\times10^{-7}$ and $2.255\times10^{-5}$, respectively. The direct references themselves agree between successive Fock cutoffs to better than $1.2\times10^{-12}$. The graph is unchanged between the two calculations. This shows that related Hamiltonians can share one compiled algebra without a new algebraic compilation.

The example also reinforces the distinction introduced in the Methods. Reusing a graph guarantees that the same declared algebra is applied to both parameter sets; it does not by itself establish that a finite operator space remains adequate after changing the model, observable, or time window.

\subsection{Large generated graphs are constructed near their final storage footprint}

Physical reuse is useful only if the graph itself can be constructed at the scales produced by the operator algebra. We therefore separate execution from the cost of compiling the graph. Complete degree-four registries with two through five boson modes provide a controlled scaling sequence: the retained operator dimension grows from 280 to 4,004, whereas the number of generated coefficient contributions grows much more rapidly.

\begin{figure*}[t!]
 \centering
 \includegraphics[width=0.90\textwidth]{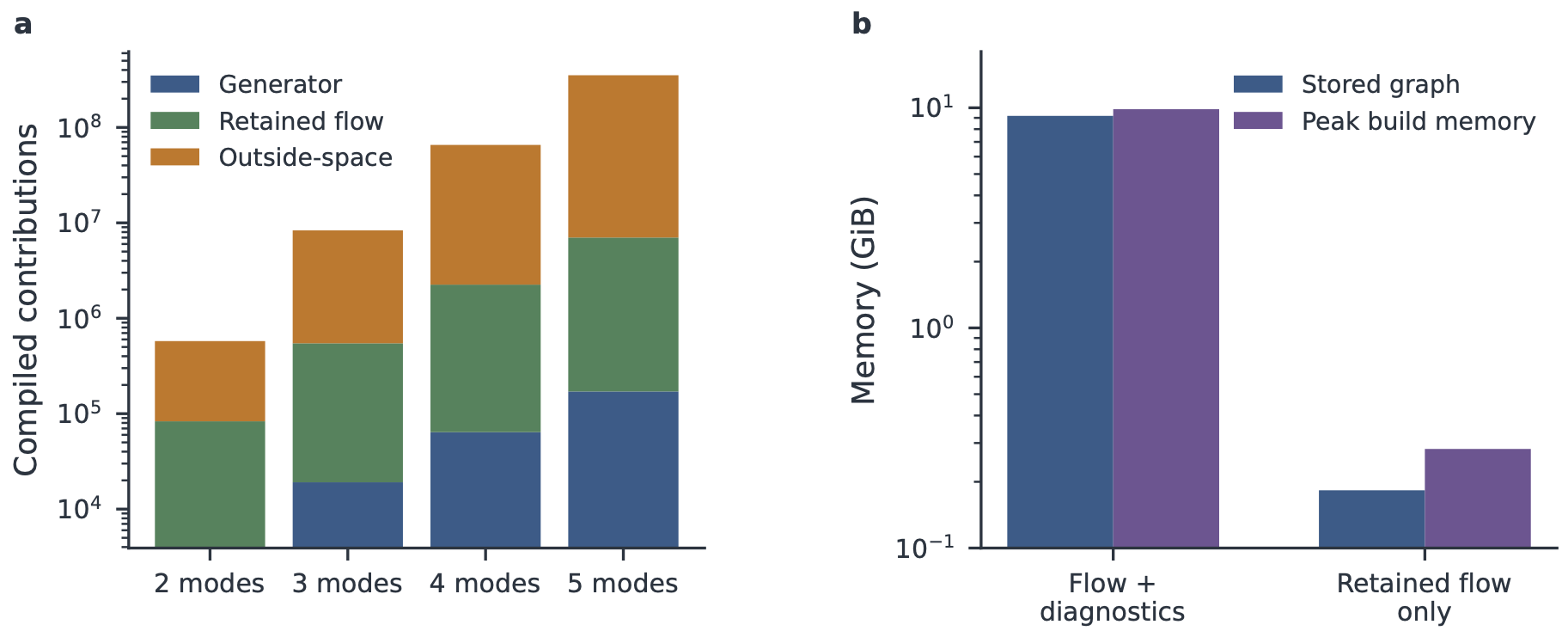}
 \caption{\label{fig:bounded-memory}
 Growth and memory footprint of the compiled Pauli--boson algebra. (a) Generator, retained-flow, and outside-retained-space contributions for complete degree-four registries with two through five boson modes. The five-mode case contains 7.017 million production contributions and 345.388 million optional outside-space records. The bars are cumulative stacks: each section starts where the preceding section ends, so the top of each colored section gives the cumulative total. (b) Final stored-graph size compared with peak memory during construction. For the diagnostic-enabled graph, 9.19~GiB of stored arrays require 9.85~GiB peak host memory (peak/final $=1.072$); the retained-flow-only graph occupies 0.183~GiB and peaks at 0.282~GiB.}
\end{figure*}

Figure~\ref{fig:bounded-memory} shows both the algebraic growth and the memory contract. For five modes, the generator and retained flow contain 7,017,036 production contributions. Recording every term generated outside the retained space adds 345,388,180 optional contributions, bringing the diagnostic-enabled total to 352,405,216. The corresponding stored graph occupies 9.19~GiB, while peak host memory during construction is 9.85~GiB, a peak-to-final ratio of 1.072. Thus a graph containing more than 350 million generated contributions is constructed with only about 7\% memory overhead relative to its final stored size in this diagnostic-enabled construction. The build takes 2.91~h.

The significance of this result is not the contribution count by itself, but the absence of a comparably large temporary graph-sized object. The compiler determines the final storage requirements before the second algebraic traversal writes contributions directly into those arrays. Consequently, construction memory remains close to the stored graph even when the generated equation set becomes very large.

The retained-flow-only graph is much smaller, occupying 0.183~GiB with a 0.282-GiB peak build memory. Removing the optional outside-space records does not, however, remove the underlying discovery work: candidate products must still be evaluated and normal ordered before the compiler can determine whether their destinations belong to the retained registry. Bounded-memory construction therefore changes the transient representation of the algebra, not its combinatorial growth. Definitions of stored graph size and peak build memory are provided in the Supporting Information.

\subsection{One stored graph supports forward, adjoint, and derivative calculations}

A persistent graph is useful only if its different numerical consumers interpret the stored algebra consistently. We therefore exercise the same compiled factors through independent CPU and GPU implementations, forward and adjoint traversals, directional derivatives, and fresh-process reload.

\begin{figure*}[t!]
 \centering
 \includegraphics[width=0.96\textwidth]{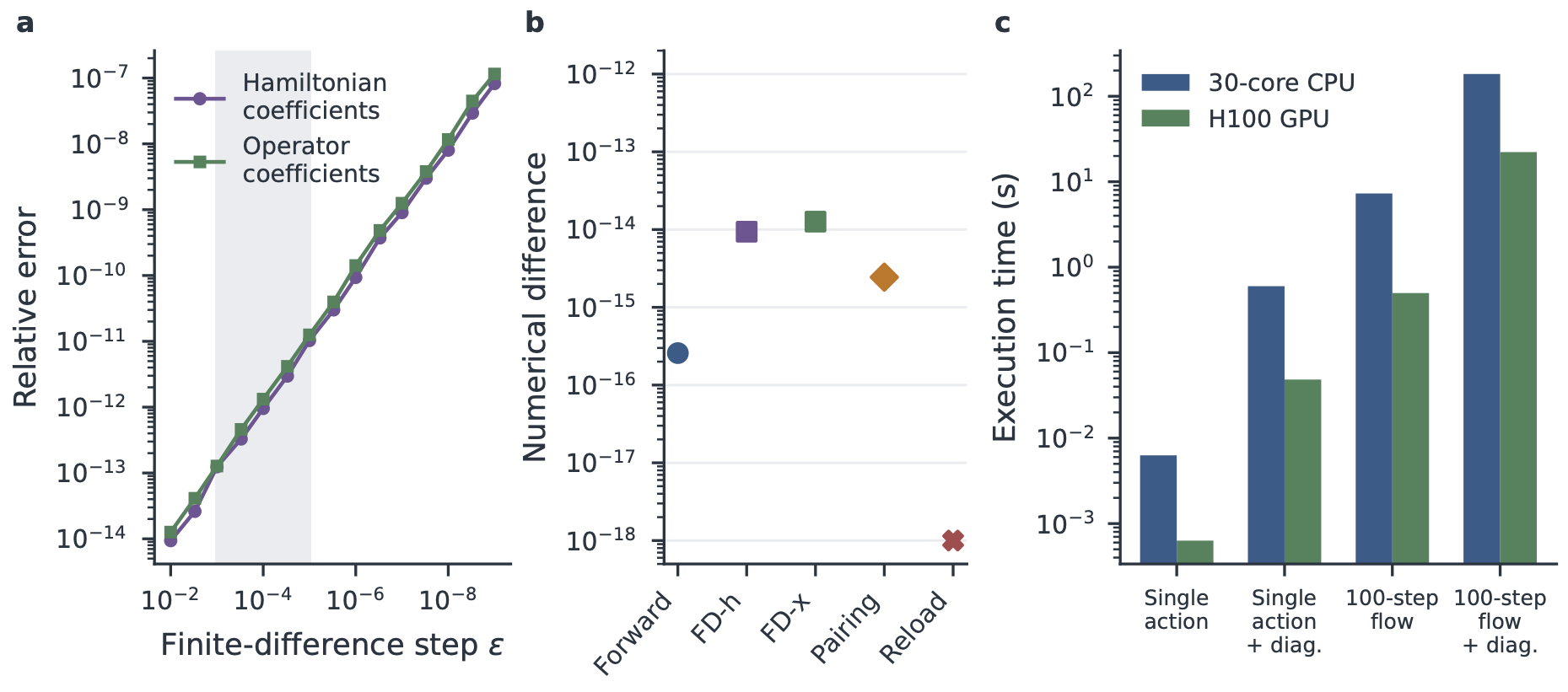}
 \caption{\label{fig:execution-consistency}
 Consistency across numerical uses of one compiled graph. (a) Directional derivatives with respect to Hamiltonian and operator coefficients are compared with centered finite differences; the shaded band marks the finite-difference validation window. (b) ``Forward'' is a relative CPU/GPU difference, ``FD-h'' and ``FD-x'' are minimum relative errors over the full scan (validation-window maxima are given in the text), and ``Pairing'' is an absolute defect. ``Reload'' denotes exact agreement and is shown at $10^{-18}$ only as a logarithmic display floor. (c) Resident execution times for a single retained-flow action and a 100-step flow, with and without optional outside-space diagnostics, on 30 physical CPU cores and one H100 GPU. Panel (c) excludes offline compilation, loading/lowering, setup, and first-call initialization.}
\end{figure*}

The derivative action is tested directly against centered finite differences (Fig.~\ref{fig:execution-consistency}a). Over the finite-difference validation window $\epsilon=10^{-3}$--$10^{-5}$, the maximum relative errors are $1.029\times10^{-11}$ for perturbations of the Hamiltonian coefficients and $1.254\times10^{-11}$ for perturbations of the operator coefficients; the minimum errors across the full scan reach $9.38\times10^{-15}$ and $1.26\times10^{-14}$, respectively. The Hamiltonian-direction test includes the induced change of the CUT generator rather than treating $L(h)$ as fixed.

Additional consistency checks are summarized in Fig.~\ref{fig:execution-consistency}b. For the five-mode benchmark, matched CPU--GPU relative differences are $2.58\times10^{-16}$ for the forward action, $2.71\times10^{-16}$ for the directional derivative, $2.95\times10^{-16}$ for the Hamiltonian adjoint derivative, and $2.54\times10^{-16}$ for the operator adjoint derivative. The absolute forward--adjoint pairing defect remains below $2.45\times10^{-15}$ in the deterministic derivative test, while the 100-step pairing drift is $2.62\times10^{-18}$. Two deterministic builds produce bitwise-identical stored graph arrays, and a newly launched process reloads the graph and reproduces every tested forward, adjoint, and derivative action exactly.

Figure~\ref{fig:execution-consistency}c shows the corresponding resident execution costs for the large five-mode graph. For a batch of eight operators, one retained-flow action takes 0.632~ms on one NVIDIA H100 GPU and 6.289~ms on 30 physical AMD EPYC 9124 CPU cores. Including all outside-space diagnostics increases those times to 48.439 and 599.488~ms. Over a 100-step flow, the retained calculation takes 0.498~s on the GPU and 7.287~s on the CPU, whereas evaluating the diagnostic at every right-hand side increases the times to 22.216 and 182.201~s.

The retained trajectories are unchanged by whether the optional outside-space records are evaluated. The timing difference therefore reflects additional diagnostic work rather than a change in the selected equations. More broadly, the agreement across processors, propagation directions, derivatives, and reload demonstrates that one compiled graph can serve several numerical consumers without separately encoding the algebra for each.

\subsection{Persistence and repeated reuse define the useful regime}

The preceding tests establish that a graph can be constructed and consumed consistently. Persistence introduces a separate practical question: when is it worthwhile to pay the one-time cost of compiling and saving the graph rather than retaining an in-memory object plan or regenerating the algebra?

A small benchmark resolves the lifecycle into compilation, reload, setup, and execution. Compilation and persistence require 0.559~s, validated reload 0.0645~s, and execution setup 0.00170~s; resident graph execution then requires $1.43\times10^{-4}$~s per retained forward action. For comparison, an independently materialized object plan takes 0.302~s to construct and $3.09\times10^{-4}$~s per action.

\begin{figure*}[t!]
 \centering
 \includegraphics[width=0.90\textwidth]{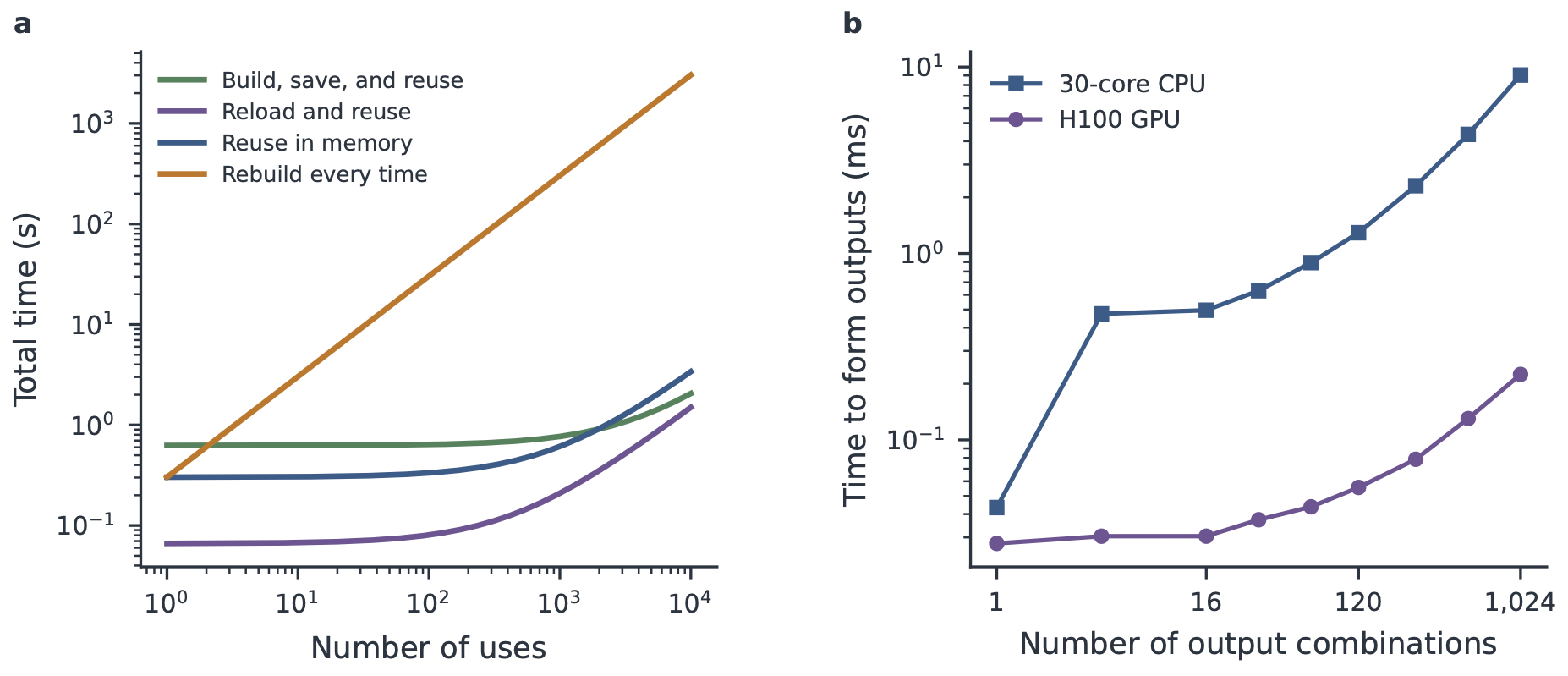}
 \caption{\label{fig:reuse-regime}
 When compile-once reuse pays. (a) Total time as a function of the number $N$ of retained forward actions for four measured lifecycle choices: build/save/reuse, reload an existing graph, retain an in-memory representation, or rebuild the algebra for every action. Loading and setup are charged once in the graph-reuse curves. Building and saving becomes favorable relative to rebuilding after three actions, while same-process in-memory reuse remains faster until 1,947 actions. (b) Time to form increasing numbers of output combinations from an already completed 120-source-operator map. CPU and GPU results agree exactly; the H100 GPU assembles 1,024 outputs in 0.224~ms.}
\end{figure*}

These measured phases define the four reuse curves in Fig.~\ref{fig:reuse-regime}a. Here, a use denotes one retained forward action under the phase definitions in the Supporting Information. Because the graph-reuse curves describe repeated actions within one lifecycle, loading and setup are charged once. If the algebra would otherwise be rebuilt for every calculation, building and saving a persistent graph recovers its one-time cost after three uses. If the object plan can instead be constructed once and retained in the same process, persistence does not break even until 1,947 uses. The value of persistence is therefore not that disk-backed execution must always outperform same-process reuse. Its distinctive benefit is that the compiled algebra becomes a validated artifact that can be reloaded and reused across process or compute-allocation boundaries without repeating symbolic generation.

Figure~\ref{fig:reuse-regime}b tests a complementary form of reuse across operators. A fixed library of 120 source operators is transformed once, after which the resulting coefficient map assembles up to 1,024 deterministic output combinations without another symbolic expansion. At 1,024 outputs, the median assembly time is 9.052~ms on the 30-core CPU and 0.224~ms on the H100 GPU, with exact agreement between the resulting coefficient arrays. This measurement isolates post-transformation reuse of a completed operator map; it does not include the generation of new physical property operators or recompilation of the transformation.

A separate comparison with a finite-Fock representation further clarifies this regime. For the present low-dimensional model, a matched 672-state finite-Fock eigentransformation remains 380--480 times faster than the compiled coefficient-space route through the tested 120-operator library. Here, matched means that both routes use the same physical Hamiltonian and source-operator library, not identical flow endpoints or output representations. In that setting, the explicit finite-dimensional representation is the natural choice. The comparison does not rank identical output representations: finite Fock produces cutoff-dependent dense matrices, whereas the operation graph acts in the selected operator-coefficient space. Its purpose is therefore to identify a representation boundary rather than a universal performance ranking.


\section{Discussion}\label{sec:discussion}

\subsection{Generated operator algebra as a persistent computational object}

The central result is that generated operator algebra can be promoted from a transient symbolic product to a persistent numerical object with a defined lifecycle: compilation, validation, storage, reload, and execution. The numerical evidence shows that this object is more than a static record of the flow equations: it survives process boundaries, accepts related physical coefficient sets, supports forward, adjoint, and derivative traversals, and is consumed consistently by CPU and GPU implementations. The novelty is therefore not the existence of fixed commutator coefficients within a finite CUT basis, but their realization as a persistent execution object with a common construction, validation, reload, and multi-consumer lifecycle.

The numerical examples probe complementary aspects of this architecture. The H/D calculation demonstrates reuse across related physical parameterizations: the coefficients change while the operator support and compiled algebra remain fixed. The large five-mode calculation instead tests the construction itself, showing that hundreds of millions of generated contributions can be compiled with peak memory close to the final stored graph. The consistency and lifecycle measurements then show that the same representation can survive reload and serve several numerical consumers. These are distinct requirements of a reusable computational representation, and separating them clarifies what each benchmark establishes.

\subsection{When compile-once reuse is useful}

Compile-once reuse is most natural when algebraic structure remains fixed while the numerical work changes. Parameter ensembles, isotope substitutions, multiple observables, derivative calculations, and repeated compute allocations all have this form. In such settings, normal ordering and equation generation need not be repeated for every coefficient vector, operand, derivative direction, or processor backend. Persistence adds a further advantage when the compiled algebra must survive process or allocation boundaries, while batched execution is useful when many observables or source operators share the same transformation.

The same architecture is not advantageous in every regime. A one-off calculation may not amortize compilation. A regular tensor contraction may already map efficiently onto established dense or sparse linear-algebra kernels. When the relevant Hilbert space is compact, an explicit matrix representation can be substantially faster than traversing a generated operator graph, as demonstrated by the finite-Fock control. Large contribution counts therefore do not constitute an advantage by themselves. The appropriate choice depends on the mathematical object required, the cost of generating it, the number and type of subsequent uses, and the structure of the competing formulation.

\subsection{Execution correctness and approximation quality are separate questions}

A compiled graph reproduces the finite projected algebra from which it was generated, but numerical agreement among independent executions does not establish that the selected operator space is a physically adequate approximation. The verification tests in the Supporting Information therefore address execution correctness: derivative and adjoint consistency, agreement between CPU and GPU implementations, deterministic rebuild and reload, and the evaluation of optional outside-space diagnostics. These checks validate the implemented finite equations, while the adequacy of the retained operator space remains a separate physical question that must be assessed against the observable, parameter regime, and time window of interest. Outside-space records should accordingly be interpreted as algebraic diagnostics rather than observable-error certificates.

This separation also suggests a natural direction for future development. Ranking omitted operators only by the magnitude of generated outside-space contributions neglects how strongly those directions influence the observable of interest. The present representation already provides two ingredients for a goal-oriented alternative: generated terms outside the retained space can be recorded, and the same compiled factors support adjoint propagation. This suggests weighting an omitted contribution by the adjoint sensitivity of a target observable, in the spirit of goal-oriented error estimation~\cite{Estep1995,Cao2004}. Such an adaptive strategy requires prospective validation and lies beyond the present work, but it provides a concrete route from algebraic residual information to observable-specific operator-space refinement.


\section{Conclusions}\label{sec:conclusions}

We have developed a compile-once numerical representation for finite Pauli--boson continuous-unitary-transformation algebra. Operator products are generated once and converted into a persistent operation graph whose stored factors can be reused as physical parameters, operators, derivative directions, and numerical backends change. A reduced HBQ example demonstrates reuse of one compiled graph across H and D parameterizations, while a separate large-scale stress test shows that a diagnostic-enabled graph containing more than 350 million generated contributions can be constructed with peak host memory only about 7\% above its final stored size.

The same compiled algebra supports forward, adjoint, derivative, and multi-operator calculations on CPU and GPU with agreement to numerical precision, and deterministic rebuild/reload tests reproduce the stored graph exactly. Lifecycle measurements further distinguish the regimes in which persistence is advantageous from those in which same-process reuse or a compact finite-dimensional representation is preferable. By separating the one-time generation of operator algebra from the numerical calculations that repeatedly consume it, the resulting architecture provides a reusable foundation for parameter studies, transformed
observables, derivative-based workflows, and heterogeneous execution, with the physical adequacy of the chosen finite operator space validated separately.


\section*{Supporting Information}

Algebraic definitions and numerical controls for the compiled Pauli--boson operation graph; the H/D model and dynamics protocol; derivative, adjoint, and fresh-process reload validation; persistent-graph contents and bounded-memory measurement definitions; CPU/GPU execution, lifecycle, and multi-operator timing protocols; and the matched finite-Fock representation comparison (PDF).

\begin{acknowledgments}
B.P. acknowledges support from the Early Career Research Program of the U.S. Department of Energy, Office of Science, under Grant No. FWP 83466. N.G. acknowledges support from a U.S. Department of Energy Scientific Discovery through Advanced Computing (SciDAC) project. B.P. acknowledges Dr. Karol Kowalski for helpful discussions.
\end{acknowledgments}

\section*{Author declarations}
\subsection*{Conflict of interest}
The authors declare no competing financial interest.

\subsection*{Data and software availability}
The source, tests, pinned environment, compiler definitions, manifests, checksums, and machine-readable evidence used for this study are maintained in the project repository. A deterministic small release case rebuilds and verifies the graph without requiring the multi-gigabyte arrays in 4- and 5-mode cases. Before submission, the rights holder will select the software license and the tagged release and archival identifiers will replace this staged statement.

\bibliography{references}

\end{document}